\documentclass[twocolumn,superscriptaddress,groupedaddress,amsmath,amssymb,aps,prl,floatfix]{revtex4-1}

\usepackage{comment}
\usepackage{graphicx}
\usepackage{dcolumn}
\usepackage{bm}

\usepackage[utf8]{inputenc}
\usepackage[T1]{fontenc}

\usepackage{braket}
\usepackage{siunitx}
\usepackage{xcolor}

\usepackage[normalem]{ulem}

\def\Rb87{^{87}\mathrm{Rb}}                             
\def\kr{k_{\mathrm r}}                             
\def\Er{E_{\mathrm r}}                             
\def\rhos{\rho^\mathrm{sf}}
\def\rhon{\rho^\mathrm{n}}

\def\ex{\mathbf{e}_x}  
\def\ey{\mathbf{e}_y}  
\def\ez{\mathbf{e}_z}  

\def\fs{f^\mathrm{sf}}  
\def\fn{f^\mathrm{n}}

\begin{document}

\title{Observation of anisotropic superfluid density in an artificial crystal}

\author{J. Tao}
\thanks{These two authors contributed equally}
\affiliation{Joint Quantum Institute, University of Maryland and National Institute of Standards and Technology, College Park, Maryland 20742, USA }
\author{M. Zhao}
\thanks{These two authors contributed equally}
\affiliation{Joint Quantum Institute, University of Maryland and National Institute of Standards and Technology, College Park, Maryland 20742, USA }
\author{I.~B.~Spielman}
\affiliation{Joint Quantum Institute, University of Maryland and National Institute of Standards and Technology, College Park, Maryland 20742, USA }
\email{ian.spielman@nist.gov}

\date{\today} 

\begin{abstract}
We experimentally and theoretically investigate the anisotropic speed of sound of an atomic superfluid (SF) Bose-Einstein condensate in a 1D optical lattice.
Because the speed of sound derives from the SF density, this implies that the SF density is itself anisotropic.
We find that the speed of sound is decreased by the optical lattice, and the SF density is concomitantly reduced.
This reduction is accompanied by the appearance of a normal fluid in the purely Bose condensed phase.
The reduction in SF density---first predicted [A.~J.~Leggett, Phys. Rev. Lett. {\bf 25} 1543--1546 (1970)] in the context of supersolidity---results from the coexistence of superfluidity and density modulations, but is agnostic about the origin of the modulations.
We additionally measure the moment of inertia of the system in a scissors mode experiment, demonstrating the existence of rotational flow. 
As such we shed light on some supersolid properties using imposed, rather than spontaneously formed, density-order.
\end{abstract}

\maketitle

Superfluidity and Bose-Einstein condensation (BEC) are deeply connected.
In dilute atomic BECs, the superfluid (SF) and condensate densities are generally equal~\cite{Ensher1996,Dalfovo1999}.
By contrast, SF $^4$He can be a nearly pure SF, with only about $14\ \%$ condensate fraction~\cite{Sears1982}, and infinite 2D Berezinskii–Kosterlitz–Thouless (BKT) SFs have no condensate at all~\cite{Berezinskii1972,Kosterlitz1973}. 
In 1970 Tony Leggett showed that supersolids---systems spontaneously forming both SF and crystalline order (i.e., density modulations)---exhibit the reverse behavior: SF density far below the condensate density~\cite{leggett1970can}.
Here we observe this effect in a nearly pure atomic BEC with artificial crystal order imprinted by an optical lattice.

The complex-valued order parameter $\phi({\bf r}) = \sqrt{\rhos} \exp[i\varphi({\bf r})]$, describing a SF with number density $\rhos$ and phase $\varphi({\bf r})$, gives rise to two hallmark SF properties: dissipationless supercurrents associated with spatial gradients in $\varphi({\bf r})$ and (Bogoliubov~\cite{Dalfovo1999}) sound described by traveling waves in $\varphi({\bf r})$.
Because supercurrents arise from phase gradients, they are locally irrotational;
in liquid $^{4}$He, the resulting non-classical rotational inertia appears below the SF transition temperature $T_{\rm c}$. 
Supersolids are more exotic systems spontaneously forming crystalline order while exhibiting SF transport properties. 
Recent experiments with dipolar BECs of Dy and Er are suggestive of these properties~\cite{tanzi2021evidence,norcia2021two}.
Leggett argued that the modulated density $\rho({\bf r})$ of a supersolid leads to an unavoidable reduction in $\rhos$, and derived an upper bound for $\rhos$~\cite{leggett1970can}.
This reduction results from the 3D density distribution, and as such is masked in tight binding descriptions such as the Bose-Hubbard model, which makes the unrelated prediction of vanishing $\rhos$ at the superfluid to Mott insulator transition~\cite{Fisher1989,greiner2002quantum}.

\begin{figure}[t]
\includegraphics{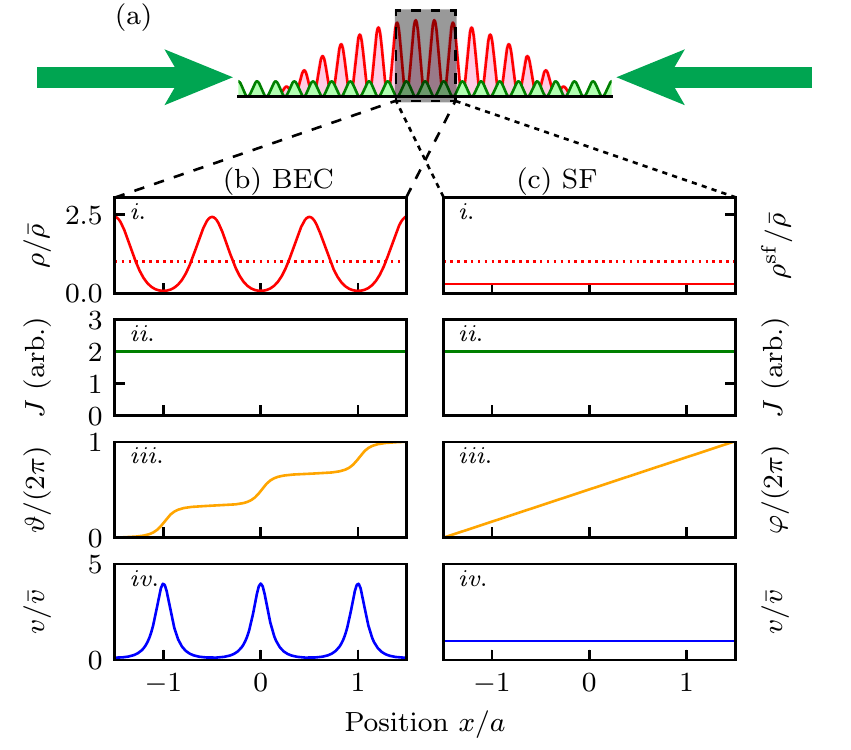}
\caption{
Concept. (a) A BEC is confined in a harmonic trap superimposed with a 1D optical lattice (along $\ex$, green), spatially modulating the condensate density (red).
The dashed and dotted lines call out a region of nominally constant mean density and the left and right columns indicate the (b) state of the condensate and (c) SF in the presence of a current.
These were computed for a $5\Er$ deep lattice and plot: {\it i}. density (red), {\it ii}. current (green), {\it iii}. phase (orange), and {\it iv}. local velocity (blue).
The red dashed line plots the mean density $\bar\rho$.
}
\label{fig:experiment}
\end{figure}

We created an artificial SF crystal by imprinting periodic density modulations into an atomic BEC using a 1D optical lattice as in Fig.~\ref{fig:experiment}(a).
While these modulations do not form spontaneously, Leggett's result still applies, making this an ideal system for understanding crystalline SFs without the added complexity of spontaneously broken symmetries.
We experimentally measured an anisotropic speed of sound via Bragg spectroscopy of the phonon mode.
This implies the existence of an effective anisotropic superfluid density---which can be expressed as a second rank tensor $\rhos_{ij}$---and we find that it saturates Leggett's bound, in agreement with Gross-Pitaveskii equation (GPE) simulations.
We also determined an associated anisotropic suppression of the moment of inertia in terms of the scissor-mode frequencies~\cite{guery1999scissors,Marago2000}. 

{\it Anisotropic superfluids}---Here we consider pure 3D BECs whose condensate mode $\psi({\bf r}) = |\psi({\bf r})| \exp[i \vartheta({\bf r})]$ is well described by the Gross-Pitaveskii equation (GPE).
An optical lattice potential $V({\bf r})=(U_0/2)\cos(2 \kr x)$ periodically modulates the condensate density $\rho({\bf r}) = |\psi({\bf r})|^2$ with unit cell (UC) size $a=\pi/\kr$ [Fig.~\ref{fig:experiment}(b)-{\it i}].
By contrast, the SF order parameter $\phi({\bf r})$ is a coarse grained quantity describing system properties on a scale $\gg a$, giving the nominally uniform density in Fig.~\ref{fig:experiment}(c)-{\it i}.

Even disregarding potential differences in $\rhos({\bf r})$ and $\rho({\bf r})$, we argue that $\phi({\bf r})$ is not simply equal to $\psi({\bf r})$ averaged over some scale large compared to $a$.
The fundamental origin of this effect can be understood by considering a 1D system of size $L$ with periodic boundary conditions in which both the condensate phase $\vartheta$ and SF phase $\varphi$ wind by an integer multiple $N$ of $2\pi$ [Fig.~\ref{fig:experiment}(b,c)-{\it ii}], yielding a metastable quantized supercurrent~\cite{fisher1973helicity}.
To satisfy the steady-state continuity equation, the microscopic current $J(x) = \rho(x) \left[\hbar \partial_x \vartheta(x) / m\right]$ must be independent of $x$ [Fig.~\ref{fig:experiment}(b)-{\it ii}], however, the periodically modulated density $\rho(x)>0$ implies the local velocity $v(x) = \hbar \partial_x \vartheta(x)/m$ has oscillatory structure and consequently $\vartheta(x)$ follows a staircase pattern  [Fig.~\ref{fig:experiment}(b)-{\it iii, iv}].

From macroscopic considerations the superfluid current is $J = \rhos \left[\hbar \partial_x \varphi(x) / m\right] = 2\pi N \hbar \rhos / (m L)$.
Equating the currents obtained from considering the condensate mode and the SF order parameter and integrating over a unit cell yields Leggett's equation~\cite{leggett1970can}
\begin{align}
\rhos &=\!a\!\left[\int_\mathrm{UC} \frac{dx}{\rho(x)}\right]^{-1}\!\!\!\!, & \!\!\text{as well as}\! &&\!\!\varphi &= \frac{1}{a} \int_\mathrm{UC}\!\!\vartheta(x) dx. \label{eq:Leggett}
\end{align}
This implies that $\rhos\leq \bar\rho$, where $\bar\rho$ is the spatial average of the condensate density over a UC, and as we discuss below the remaining density $\rhon=\bar\rho - \rhos$ behaves as a pseudo-normal fluid. 
In the more general context where the GPE is inapplicable, the Leggett expression for $\rhos$ is an upper bound for the SF density in systems with crystalline order~\cite{leggett1970can}.

In a 3D system, the current $J_i=\rhos_{ij}\left[\hbar \partial_j \varphi/m\right]$ derives from a SF density tensor.
For systems with rectangular symmetry~\footnote{It also necessary that the condensate mode be of a separable form~\cite{Leggett1998}.} $\rhos_{ij}$ is diagonal, and the analogs to Eq.~\eqref{eq:Leggett} for each of the three elements use a 1D density integrated along the transverse directions. 
In our experiments this implies that the superfluid density is only reduced along the direction of the optical lattice, so $\rhos_{yy} = \rhos_{zz} = \bar\rho$.

{\it Experiment}---We used $^{87}\mathrm{Rb}$ BECs with $N\approx2\times10^5$ atoms in the $\ket{F=1, m_F=1}$ hyperfine ground state.
A $1064\ \mathrm{nm}$ trapping laser with an elliptical cross-section, traveling along $\ex$ provided strong vertical confinement with frequency $\omega_z/(2\pi) = 220\ \mathrm{Hz}$; the in-plane frequencies, from $\omega_{x,y}/(2\pi)=(34,51)\ \mathrm{Hz}$ to $(56,36)\ \mathrm{Hz}$, were optimized for our different experiments.
We created a 1D optical lattice using a retro-reflected $\lambda = 532\ \mathrm{nm}$ laser traveling along $\ex$, giving an $a=266\ \mathrm{nm}$ lattice period, comparable to the $\xi = 170(20) \ \mathrm{nm}$ minimum healing length.
The optical lattice was linearly ramped on in $100\ \mathrm{ms}$ to a final depth $\leq 10\ \Er$, with single photon recoil energy and momentum $\Er=\hbar^2 \kr^2 / (2 m)$, and $\hbar\kr = 2\pi\hbar/\lambda$ respectively~\footnote{We calibrated the lattice depth $U_0$ by suddenly applying the lattice potential and fitting the resulting Kaptiza-Dirac scattering~\cite{Huckans2009a}.}.
For Bragg experiments the final state was measured using resonant absorption imaging after a  $15\ \mathrm{ms}$ time of flight (TOF); scissors mode measurements were performed in-situ using partial transfer absorption imaging~\cite{Ramanathan2012}.

\begin{figure*}[tb]
\includegraphics{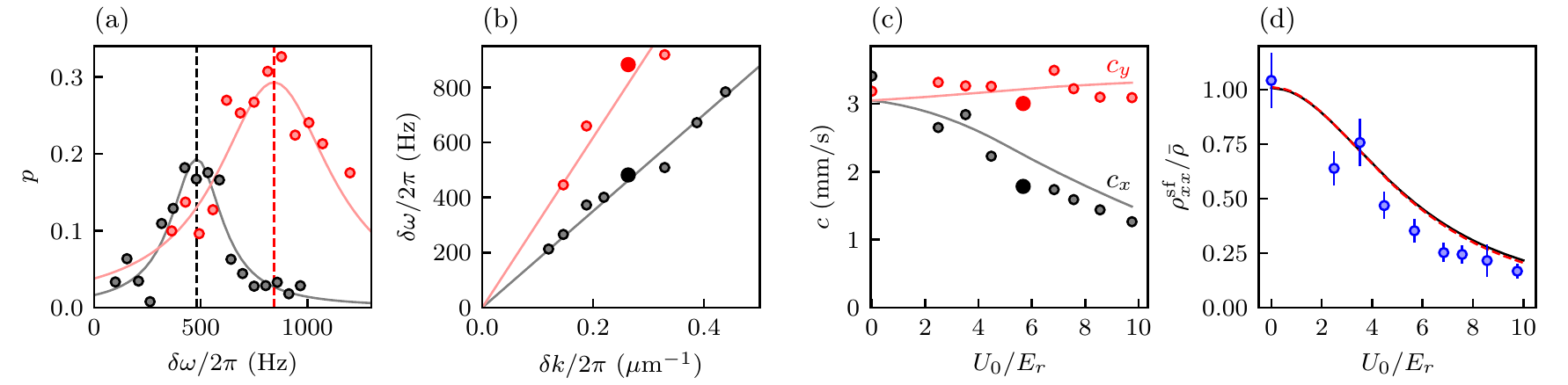}
\caption{\label{fig:sound_speed} 
Bragg spectroscopy. 
Black and red symbols mark excitations created along $\ex$ and $\ey$ respectively. 
(a) Transferred population fraction $p$ as a function of frequency difference $\delta\omega$ with wavevetor $\delta k / 2\pi = 0.26\ \mu{\rm m}^{-1}$ and lattice depth $U_0 = 5.7 \Er$.
The solid curve is a Lorentzian fit giving the resonance frequency marked by the vertical dashed line.
(b) Phonon dispersion obtained from Bragg spectra.  
The bold symbols resulted from (a) and the linear fit (with zero intercept) gives the speed of sound.
(c) Anisotropic speed of sound.
The bold symbols are derived from (b) and the solid curves are from BdG simulations (no free parameters~\cite{SeeSM}).
(d) SF density obtained from speed of sound measurements (blue markers, error bars mark single-sigma statistical uncertainties). 
We compare with two models: the red dashed curve plots a homogeneous gas BdG calculation, and the solid black curve plots the result of Eq.~\eqref{eq:Leggett}.
The simulations used our calibrated experimental parameters.
}
\end{figure*}

{\it Anisotropic speed of sound}---The speed of sound for diagonal $\rhos_{ij}$ is predicted to result from $c_i^2 = \fs_{ii} / (\kappa m)$ in terms of the superfluid fractions $\fs_{ii}=\rhos_{ij}/\bar\rho$, the compressibility $\kappa = \bar\rho^{-1} \left(\partial \bar\rho / \partial \mu \right)$, and the chemical potential $\mu$.
This reduces to the well-known value $c^2 = \mu / m$ for an isotropic homogeneous system (See \cite{SeeSM} for the full dispersion beyond the linear approximation).
The sound speed ratio
\begin{align}\label{SupFraction}
\frac{c_x^2}{c_y^2} &= \frac{\rhos_{xx}}{\rhos_{yy}} = \fs_{xx},
\end{align}
provides direct access to the different components of the superfluid density [see \cite{SeeSM} for a Josephson sum rule~\cite{clark2013mathematical} argument].
Because the density is $y$-independent, Eq.~\eqref{eq:Leggett} implies $\rhos_{yy}=\bar\rho$.

We Bragg-scattered the BEC off a weak sinusoidal potential with reciprocal lattice vector $\delta k$ slowly moving with velocity $v$ by patterning a laser beam with a digital micro-mirror device (DMD~\cite{ha2015roton}) and measured the scattered fraction $p$.
This results from what are effectively two interfering laser beams driving two-photon transitions with difference-wavevector $\delta k$ and angular frequency $\delta \omega = \delta k\ v$.
We applied this potential for $\approx\SI{5}{\milli\second}$.
Bragg transitions ensued when the difference energy and momentum were resonant with the BEC's Bogoliubov dispersion, and Fig.~\ref{fig:sound_speed}(a) shows data in the linear regime.
The width of this spectral feature is limited by our BEC's inhomogeneous density profile; the resonance (vertical dashed line) obtained from a Lorentzian fit (solid curve) therefore reflects an average speed of sound~\footnote{In high-elongated quasi-1D BECs, the longitudinal speed of sound is reduced by a factor of $\sqrt{2}$ from $\sqrt{\mu/m}$.
We expect a related reduction from our tight confinement along $\ez$, but for the Bragg spectra to exhibit inhomogeneous broadening from the nearly isotropic Thomas-Fermi profile in the $\ex$-$\ey$ plane.}.

A series of such fits lead to phonon dispersion relations with Bragg-lattice period from $2.25\ \mu{\rm m}$ to $8.5\ \mu{\rm m}$.
Representative dispersions taken along $\ex$ and $\ey$ are shown in  Fig.~\ref{fig:sound_speed}(b), and we obtain the phonon speed of sound using linear fits. 
Figure~\ref{fig:sound_speed}(c) summarizes these data showing the speed of sound decreasing along the lattice direction $\ex$, but slightly increasing along $\ey$ (resulting from the increased atomic density in the individual lattice sites).
Finally Fig.~\ref{fig:sound_speed}(d) shows our main result: the normalized superfluid density obtained from these data using Eq.~\eqref{SupFraction} decreases as a function of $U_0$.

We compared these data to GPE simulations in two ways, we: (1) used the Bogoliubov-de Gennes (BdG) equations to obtain $c_x$ and $c_y$ and (2) directly evaluated Eq.~\eqref{eq:Leggett} from the GPE ground state density.
The solid curves in Fig.~\ref{fig:sound_speed}(c) plot the sound speed obtained from solving the 1D BdG~\footnote{To make $k$ a good quantum number we modeled untrapped systems with periodic boundary conditions.
The chemical potential was selected to give the observed $3\ {\rm mm}/{\rm s}$ speed of sound without the lattice present.}, and the red dashed curve in (d) is the ratio of these speeds.
To compare with Leggett's prediction, we found the ground state of the 2D GPE for our experimental parameters and evaluated Eq.~\eqref{eq:Leggett} throughout our inhomogeneous system.
The black curve in Fig.~\ref{fig:sound_speed} plots the resulting weighted average.
Remarkably the BdG results are in near-perfect agreement with Leggett's expression.

\begin{figure}[b!]
\includegraphics{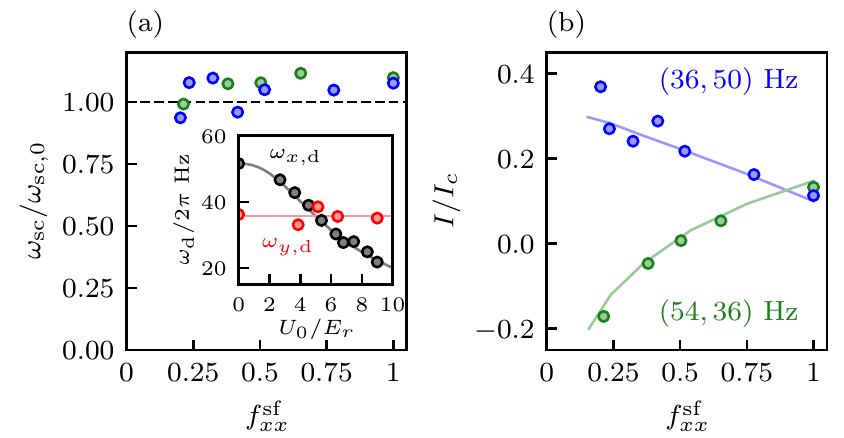}
\caption{Moment of inertia from scissors mode. 
(a-inset) Measured dipole mode frequencies (markers) along with fits (curves) where the frequency at $U_0$ is the only free parameter for each curve. 
(a) Scissors mode frequency.
Blue and green correspond to $U=0$ trap frequencies $(34,51)\ {\rm Hz}$ and $(54, 36)\ {\rm Hz}$ respectively.
(b) Moment of inertia in units of $I_{\rm c}$.
Symbols are the data computed as described in the text, and the solid curves are GPE predictions using
$I=\partial_{\Omega}{\langle L_z \rangle}$, with angular frequency $\Omega$.
} \label{fig:collective_modes} 
\end{figure}

{\it Scissors mode}---The single-valued nature of the SF order parameter greatly affects rotational properties such as the moment of inertia $I$.
For highly anisotropic traps, the scissors mode~\cite{guery1999scissors} describes a fixed density distribution pivoting by a small angle $\theta$ about the trap center with frequency $\omega_{\rm sc}$.
Scissors mode experiments are reminiscent of torsional balance experiments in $^4{\rm He}$~\cite{Andronikashvili1946} which give access to the non-classical rotational inertia~\cite{leggett1970can}.

It is suggestive to quantify these dynamics in terms of the Lagrangian $L = I \dot\theta^2/2 - V(\theta)$, for moment of inertia $I$ and potential energy $V(\theta)$.
For small $\theta$ the potential can be expanded as $V(\theta)\approx I \omega_{\rm sc}^2 \theta^2 /2$ with
\begin{align}
\frac{I}{I_{\rm c}} &= \frac{(\omega_x^2 - \omega_y^2)^2}{ \omega_{\rm sc}^2 (\omega_x^2 + \omega_y^2)}\label{eq:I_over_Ic}
\end{align}
in terms of the classical moment of inertia $I_{\rm c}$ and in agreement with Ref.~\cite{guery1999scissors} for isotropic superfluids.
Although this interpretation is highly intuitive, it does not survive careful consideration.
The anisotropic superfluid density couples radial and azimuthal flow and as a result a single parameter Lagrangian is insufficient to describe rotational dynamics.

Instead the superfluid hydrodynamic equations predict a moment of inertia scaled by a factor of $(\fs_{xx}\omega_x^2- \fs_{yy}\omega_y^2) / (\omega_{x}^2- \omega_{y}^2)$ (see \cite{SeeSM}) compared to Eq.~\eqref{eq:I_over_Ic}.
Therefore we expect $\omega_{\rm sc}$, in conjunction with the superfluid density will give $I/I_c$ as a function of lattice depth.

The inset to Fig.~\ref{fig:collective_modes}(a) plots the dipole mode frequencies $\omega_{x,{\rm d}}$ and $\omega_{y,{\rm d}}$ for a trap with frequencies $(54,36)\ {\rm Hz}$.
The frequency reduction is also related to $\rhos$ via $\fs = (\omega_{x,{\rm d}}/\omega_x)^2$ along the lattice direction~\cite{SeeSM}.
This ratio can also be expressed in terms of an increased effective mass that converges to the predictions of single-particle band structure~\cite{Jimenez-Garcia2013} when the lattice period falls below the healing length; in our case the value  computed perturbatively from the GPE differs by about $20\ \%$ from the band structure prediction.
The result of this modeling is shown by the solid curves.

We excited the scissors mode using our DMD to tilt the harmonic potential by $50$ to $140~{\rm mrad}$ for $\approx 1\ {\rm ms}$ (shorter than the trap periods) and let the BEC evolve in the original trap for a variable time. 
We measured the resulting dynamics in-situ and extracted the angle by fitting the resulting density profile to a rotated Gaussian.
Figure~\ref{fig:collective_modes}(a) shows scissor mode frequency normalized to the expected frequency~\cite{pitaevskii2016bose} of $\omega_{{\rm sc},0}^2 = \fs_{xx}\omega_x^2 + \fs_{yy}\omega_y^2$ for a trap elongated either along $\ex$ [with frequencies $(56,36)\ {\rm Hz}$, blue] or along $\ey$ [with frequencies $(36,50)\ {\rm Hz}$, green].
In both cases $\omega_{\rm sc}$ is about $5\ \%$ in excess of the simple prediction, perhaps from finite temperature or anharmonicities in the ODT.

We combine these observations in Fig.~\ref{fig:collective_modes}(b) to obtain $I/I_{\rm c}$; the data (symbols) and our 2D GPE simulations (curves) are in agreement.
For traps elongated along $\ex$ (green) $I/I_c$ unexpectedly changes sign when $\omega_{x,{\rm d}}=\omega_{y,{\rm d}}$.
To understand the physical origin of this effect we now turn our attention to rotating systems.

\begin{figure}
\includegraphics{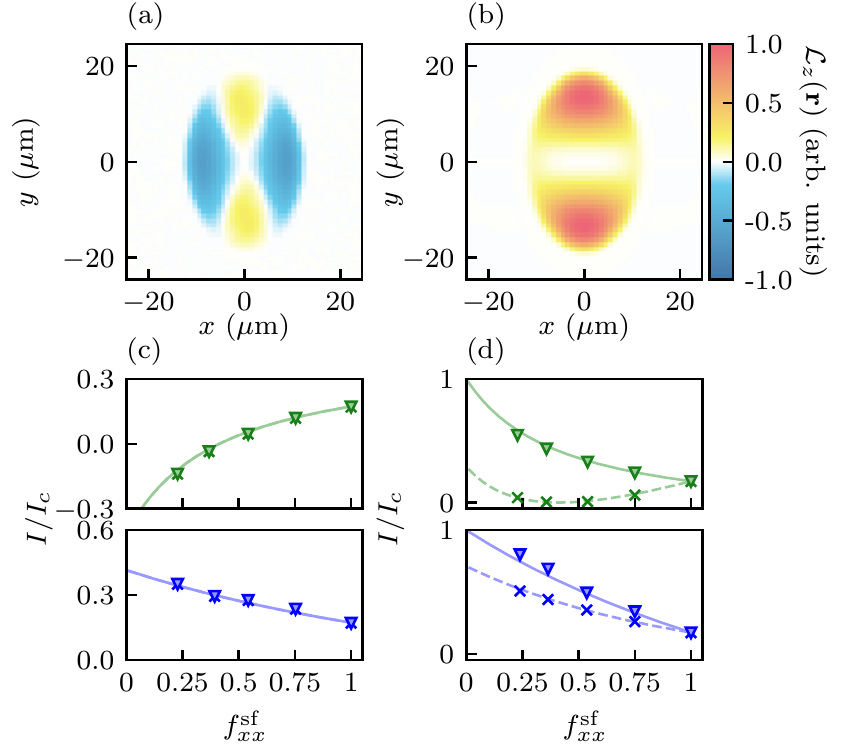}
\caption{\label{fig:NCRI_rot} 
Moment of inertia in rotating systems computed using 2D GPE simulations.
The left column indicates simulations in which the lattice is static while in the right column the lattice co-rotates with the confining potential.
(a,b) Angular momentum density for trap frequencies $2\pi\times$(56,36) and $U_0=10\Er$.
(c, d) Total momentum of inertia in traps with frequencies $2\pi\times$(56,36) (top, green) and $2\pi\times(36,56)\ {\rm Hz}$ (bottom, blue).
Dashed curves plot $I^{\rm sf} / I_c$ and the solid curve plots $I/I_c$.
}
\end{figure}

{\it Rotation}---Thus far we focused exclusively on the superfluid density, while avoiding questions about any associated normal fluid. 
We can deduce the existence of a normal fluid component by considering two thought experiments in a 1D ring geometry (with radius $R$) and quantify both in terms of the resulting angular momentum~\footnote{These arguments are not limited to phase windings that are multiples of $2\pi$.}.
In case (i), we consider an Aharonov-Bohm geometry and slowly thread the ring with a single quanta of magnetic flux (see Ref.~\cite{Cooper2010} for an artificial gauge field proposal). 
The process is equivalent to imprinting a $2\pi$ phase winding (of the type discussed on page 1), giving angular velocity $\Omega = \hbar / (m R^2)$ and angular momentum $L_z/\hbar = 2 \pi R \rhos$.
In case (ii), we consider a complimentary experiment in which the lattice is very slowly accelerated to a final angular velocity $\Omega$; this is best understood by transforming into the frame co-rotating with the lattice.
This leads to a lab frame angular momentum $L_z/\hbar = 2 \pi R (\bar \rho - \rhos)$ which we interpret as resulting from the normal fluid co-moving with the lattice.

With this insight we extended our 2D numerical simulations to analogous cases for rotating harmonically trapped systems where : (i) the lattice is static in the lab frame (as in scissors mode experiments) or (ii) it co-rotates with the confining potential.
In both cases we use the coarse graining defined in Eq.~\eqref{eq:Leggett} to obtain the superfluid density and phase.
In this way we compute the total moment of inertia $I$ from $\psi({\bf r},t)$, the superfluid component $I^{\rm sf}$ from $\phi({\bf r},t)$, and we define the normal component as the difference $I^{\rm n} = I - I^{\rm sf}$.

Case (i): as in our 1D thought experiment only the SF component responds.  
Then although $\nabla\varphi$ is manifestly irrotational, because $\rhos_{xx}\neq\rhos_{yy}$ the superfluid current can be rotational.
In this case, the relative magnitude of the co- and counter-rotating contributions vary with the lattice depth, leading to regions of negative angular momentum density ${\mathcal L}({\bf r})$ along the BEC's semi-minor axis [Fig.~\ref{fig:NCRI_rot}(a)]. 
The superfluid moment of inertia computed from these simulations [Fig.~\ref{fig:NCRI_rot}(c)] is in full agreement with the scissor mode simulation, and as expected for a static lattice $I^{\rm sf} = I$ (no normal flow).

When the lattice is along the semi-minor axis, as pictured in (a) and the green curve in (c), the counter-rotating contribution increases with $U_0$, until the dipole mode frequencies along $\ex$ and $\ey$ invert, after which point, $I/I_{\rm c}$ becomes negative.
The reverse is the case when the lattice is along the semi-major axis and $I/I_c$ increases monotonically.
This novel observation confirms the negative kinetic energy resulting from $\dot\theta$. 

Case (ii): In contrast, the angular momentum density is strictly positive [Fig.~\ref{fig:NCRI_rot}(b)] for both lattice orientations and $I/I_c$ increases with lattice depth [Fig.~\ref{fig:NCRI_rot}(d)].
In this case the normal fluid to co-rotate with the trap giving the current $J^n = (-\rhon_{xx} y, \rhon_{yy} x) \dot{\theta}
$.
The total $I/I_c$ is then the sum of the superfluid~\cite{SeeSM} and normal contribution
\begin{align}
    \label{eq:NCRI_rot}
\frac{I}{I_c}&=
 \frac{(\fs_{xx}\omega_{x}^2- \fs_{yy}\omega_{y}^2)^2}{(\fs_{xx}\omega_{x}^2+ \fs_{yy}\omega_{y}^2)(\omega_x^{2}+\omega_y^{2} ) }+ \frac{\fn_{xx} \omega_x^2 +\fn_{yy} \omega_y^2 }{\omega_x^2+\omega_y^2}.
\end{align}
This result, along with our 2D GPE simulations, are plotted in Fig.~\ref{fig:NCRI_rot}(d).
The dashed curve plots the superfluid contribution to $I^{\rm sf}/I_c$ in agreement with the coarse-grained GPE (crosses).
The solid curve and the triangles plot the corresponding total moment of inertia, in excess of the SF contribution.
This implies the appearance of normal fluid flow.

This agreement confirms that the superfluid contribution derives from  gradients of the coarse-grained phase $\varphi$, while the normal contribution stems from variations of $\vartheta$ within each lattice site.

{\it Discussion and outlook}---Our inability to obtain $I/I_{\rm c}$ from scissors mode measurements without detailed modeling reinforces similar conclusions in dipolar gases~\cite{Roccuzzo2022}.
In both cases the simple argument fails because $\dot\theta$ couples to more internal degrees of freedom than $L_z$ alone. 
In this context Ref.~\cite{Roccuzzo2022} concluded that the scissors mode {\it does} yield the moment of inertia when 1D density modulations comove with the oscillatory motion: this is consistent with our findings comparing motion in static and rotating lattices.
Our GPE simulations indicate that the analytical relations generalize to lattices with period in excess of the healing length.

Although we conclude that a normal fluid exists, it is inseparable from the optical lattice and lacks any internal dynamics of its own, i.e., it is not described by a dynamical equation of motion.
In contrast, both the superstripe phase in spin-orbit coupled BECs~\cite{Lin2011,Li2017,Li2013,zhang2016superfluid,Martone2021} and supersolid phases of dipolar gases~\cite{Tanzi2019,Bottcher2019,Chomaz2019}, support dynamical density modulations.
Leggett's expression applies to both of these systems implying a reduced superfluid density, which in this case could exhibit dynamics, as expected for a system described by a two-fluid model~\cite{guery1999scissors,Marago2000}.

This leaves open questions regarding nature the normal fluid of spin-orbit coupled systems where an interplay between single-particle physics and interactions govern supersolid-like properties.
In addition, $\rhos$ is expected to be reduced outside of the superstripe phase~\cite{zhang2016superfluid,Martone2021} where the density is uniform (making Leggett's expression inapplicable), but the BEC's spin vector is spatially periodic.


\begin{acknowledgments}
{\it Note:} During the early stages of manuscript preparation we become aware of a related work, using a long period 1D lattice applied to a homogeneously confined 2D BEC.

The authors thank S.~Stringari for suggesting this line of investigation and to both S.~Stringari and S.~Roccuzzo for stimulating discussions.
In addition W.~D.~Phillips and S.~Mukherjee carefully read the manuscript.
This work was partially supported by the National Institute of Standards and Technology, and the National Science Foundation through the Physics Frontier Center at the Joint Quantum Institute (PHY-1430094) and the Quantum Leap Challenge Institute for Robust Quantum Simulation (OMA-2120757).
\end{acknowledgments}

\bibliography{main}

\end{document}


\title{Supplementary material}

\author{J. Tao}
\affiliation{Joint Quantum Institute, University of Maryland and National Institute of Standards and Technology, College Park, Maryland 20742, USA }
\author{M. Zhao}
\affiliation{Joint Quantum Institute, University of Maryland and National Institute of Standards and Technology, College Park, Maryland 20742, USA }
\author{I.~B.~Spielman}
\affiliation{Joint Quantum Institute, University of Maryland and National Institute of Standards and Technology, College Park, Maryland 20742, USA }
\email{ian.spielman@nist.gov}

\date{\today} 

\maketitle

In this supplementary material, we derive the relationship between superfluidity and elementary excitation properties. We also derive a hydrodynamics model to analyze the angle rotation of Bose-Einstein condensates in harmonic trap in the small period lattice case.

\section{Superfluid sum-rule and sound velocity}
The superfluid density is explicitly related to Green function of the many-body Hamiltonian via  the Josephson sum-rule~\cite{clark2013mathematical}.
Here we derive the sum-rule and relate it to the measurable quantities of the anisotropic sound velocities.

We consider the variation of the condensate wavefunction $\psi(x)=\sqrt{|\psi|} \exp[i\phi({\bf r})]$ from a perturbation Hamiltonian $H'=-\int d\bf r \psi(\bf r)\xi(\bf r)$, where $\xi(\bf r)=\xi \exp{i (k\cdot r-\omega t)}$.
According to linear response theory
\begin{equation}
    \delta\langle\psi(\mathbf{r})\rangle=e^{i k\cdot r-i \omega t} \xi \int_{-\infty}^{\infty} d \omega' \frac{\mathbf{A}(\mathbf{k}, \omega')}{\omega+i\eta-\omega'},
\end{equation}
where 
$
A(\boldsymbol{k},\omega)=\int d \mathbf{r} e^{-i \boldsymbol{k} \cdot\left(\boldsymbol{r}-\boldsymbol{r}^{\prime}\right)} \int_{-\infty}^{\infty} d t e^{i \omega\left(t-t^{\prime}\right)}\left\langle\left[\psi(\boldsymbol{r} ,t), \psi^{\dagger}\left(\boldsymbol{r}^{\prime}, t^{\prime}\right)\right]\right\rangle
$
is the spectral density, and is related to the retarded one-body Green function
$$
A(\mathbf{k}, \omega) = i \hbar \int d\left(\mathbf{r}-\mathbf{r}^{\prime}\right) e^{-i \mathbf{k}\left(\mathbf{r}-\mathbf{r}^{\prime}\right)} \int_{-\infty}^{\infty} d\left(t-t^{\prime}\right) e^{i \omega\left(t-t^{\prime}\right)} G^{\mathrm{ret}}\left(\mathbf{r}, t ; \mathbf{r}^{\prime}, t^{\prime}\right)
.$$
Similarly, the variation of the current operator $\bf J( r)=\rho v$ is
\begin{equation}\label{eq:J}
     \delta \langle \mathbf{J}(r)\rangle=e^{i \mathbf{k}\cdot \mathbf{r}-i \omega t} \xi \int d \omega^{\prime} \frac{\mathbf{\Gamma}\left(\mathbf{k}, \omega^{\prime}\right)}{\omega+i \eta-\omega^{\prime}} 
\end{equation}
with
$
\Gamma(k, \omega)=\int d \mathbf{r} d t e^{-i \omega\left(t-t^{\prime}\right)+i k\left(r-r^{\prime}\right)}\langle\left[J(r, t), \psi^\dagger\left(r^{\prime}, t^{\prime}\right)\right]\rangle
.$
In the next we first take $\omega\rightarrow0$ limit, and then take $\mathbf{k}\rightarrow0$.

We want to relate $\bf \delta \langle J(r)\rangle$ to $\delta\langle\psi(\bf r)\rangle$, as to define the superfluid density. To do that, one observes from the continuity equation $-i \mathbf{k} \cdot \mathbf{J}=\partial_{t} \rho$ so that
$$
\mathbf{J}(\mathbf{r},t)=\int d \mathbf{k} \mathbf{J}(\mathbf{k},t) e^{i k \cdot r}=\int d \mathbf{k} d \mathbf{r}^{\prime} e^{i k\left(r-r^{\prime}\right)} \frac{i \vec{k}}{k^{2}} \partial_t \rho(r',t)
.$$
Plug it back to equation~\ref{eq:J}, and after some algebra
\begin{equation}
    \begin{aligned}
        \Gamma(\mathbf{k},\omega) &= \int d \mathbf{r} d t e^{-i \omega\left(t-t^{\prime}\right)+i k\left(r-r^{\prime}\right)} \int d \mathbf{k}^{\prime} d \mathbf{r}^{\prime \prime} e^{i k^{\prime}\left(r-r^{\prime \prime}\right)} \frac{i \vec{k'}(-i \omega)}{k'^2}\langle \left[\rho\left(r^{\prime \prime},t\right), \psi^{\dagger}\left( r^{\prime},t'\right)\right]\rangle\\
        &=\int d t e^{-i \omega\left(t-t^{\prime}\right)} \int d \mathbf{r}^{\prime \prime} e^{i k\left(r^{\prime \prime}-r^{\prime}\right)} \frac{\mathbf{k} \omega}{k^{2}}\langle\left[\rho\left(r^{\prime \prime} t\right), \psi^{\dagger}\left(r^{\prime} t^{\prime}\right)\right]\rangle
    \end{aligned}
\end{equation}
Hence
\begin{equation}
    \begin{aligned}
        \delta \langle \mathbf{J}(\mathbf{r},t) \rangle &=\int d \omega^{\prime} \frac{\mathbf{\Gamma}\left(k, \omega^{\prime}\right)}{-\omega^{\prime}} e^{i \mathbf{k} r} \xi\\
        &=-\int d \omega e^{i \mathbf{k} r} \xi \int d t e^{-i \omega\left(t-t^{\prime}\right)} \int d \mathbf{r} e^{i k\cdot (r-r')} \frac{\vec{k}}{k^{2}} \langle\left[\rho(r, t), \psi^{\dagger}\left(r' ,t^{\prime}\right)\right]\rangle\\
        &=-e^{i \mathbf{k} r} \xi \int d \mathbf{r} e^{i \mathbf{k} \cdot\left(r-r^{\prime}\right)} \frac{\vec{k}}{k^{2}}\left\langle\left[\rho(r t), \psi^{\dagger}\left(r^{\prime} t\right)\right]\right\rangle\\
        &=-e^{i k\cdot r} \xi \frac{k}{k^{2}}\left\langle\psi^{\dagger}(r,t)\right\rangle
    \end{aligned}
\end{equation}
In the last equality we used
$\left[\rho(r), \psi^{\dagger}\left(r^{\prime}\right)\right]=\left[\psi^{\dagger}(r) \psi(r), \psi^{\dagger}\left(r^{\prime}\right)\right]=\psi^{\dagger}(r) \delta\left(r-r^{\prime}\right)$.

For a superfluid system,
$\langle\psi(\mathbf{r})\rangle=\psi_0 + \delta\langle\psi(\mathbf{r})\rangle = e^{i \theta(\mathbf{r})} \psi_0$
where we have defined the superfluid phase $\theta(\mathbf{r})$. 
With the periodic modulation of density introduced by the optical lattice, we take 
$\psi(\mathbf{r}) \rightarrow \bar{\psi}(\mathbf{r}) = a_x a_y a_z \int_\mathrm{UC} d \mathbf{r} \psi(\mathbf{r})$.\cite{muller2015josephson}
Now we have
$$
\begin{aligned}
&\delta\langle\bar{\mathbf{J}}\rangle=-e^{i \mathbf{k} \cdot r} \xi \frac{\mathbf{k}}{k^{2}}\left\langle\bar{\psi}^{\dagger}\right\rangle \\
&\delta\langle\bar{\psi}\rangle=-e^{i \mathbf{k} \cdot r} \xi \int d \omega \frac{A(k, \omega)}{\omega}\\
&\frac{m}{\hbar} \delta\langle\bar{\mathbf{J}}\rangle=  \rho^{s f} \nabla \theta=  \rho^{s f} \nabla \frac{\delta\langle\bar{\psi}\rangle}{i \bar{\psi}_{0}}= \rho_{\hat{k}}^{s f} \mathbf{k} \frac{\delta\langle\bar{\psi}\rangle}{\bar{\psi}_0}
\end{aligned}
$$
and take $\mathbf{k}\rightarrow 0$
\begin{equation}
    \rho_{\hat{k}}^{s f}=m  \lim _{\mathbf{k} \rightarrow 0} \frac{\bar{\psi}_{0}^{*} \bar{\psi}_{0}}{k^{2} \int_{-\infty}^{+\infty} d \omega \frac{\mathbf{A}(k, \omega)}{\omega}}
\end{equation}
where $\rho_{\hat{k}}^{s f}=\rho_{ij}^{s f} \mathbf{\hat{k}}_j$. 
For homogeneous weakly interacting Bose gas at $T=0$, the Bogoliubov theory gives 
$
\mathbf{A}(\mathbf{k}, \omega) = \frac{mc}{2k} [\delta(\omega-ck)-\delta(\omega+ck)]
$
as $k\rightarrow0$, which leads to $\rho_{\hat{k}}^{s f}=|\psi_0|^2$. For anisotropic systems, the poles of $\mathbf{A}(\mathbf{k}, \omega)$ i.e. the sound velocities along different directions will determine the anisotropic superfluid density.

\begin{figure}
\includegraphics{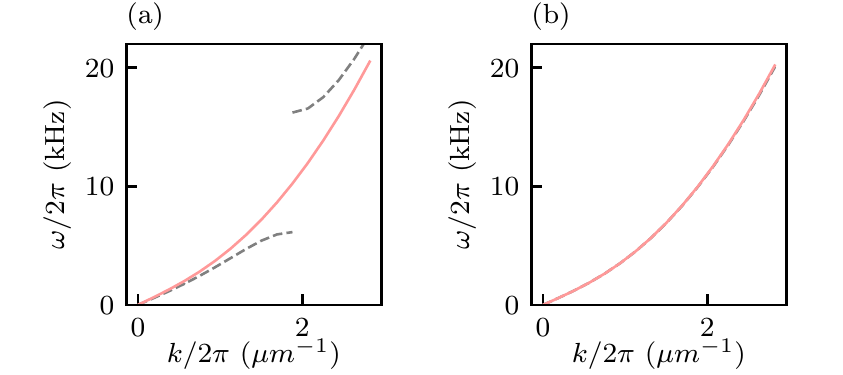}
\caption{\label{fig:Fig1_SM} 
Modification of the phonon spectrum by the $a=266$~nm optical lattice via BdG calculation. Dashed black and solid red curves mark excitations created along $\ex$ and $\ey$ respectively. (a) At $U_0=3 E_r$, (b) without the lattice.
}
\end{figure}

The modification of the phonon spectrum by the presence of a shallow optical lattice is numerically computed through Bogoliubov-de-Gennes equations applied upon the mean-field ground state found by a 2D GPE imaginary time simulation. The lattice changes the sound velocity ratio while opening a gap at the Brillouin zone edge, as in Fig.~\ref{fig:Fig1_SM}.

\section{Hydrodynamics theory for Bose-Einstein condensates.}

\subsection{Coarse-grained hydrodynamics}

For our BEC in an optical lattice where the healing length is larger or comparable to the lattice period, the long wavelength dynamics can be well described by the coarse-grained hydrodynamics picture where the microdynamics on the lattice scale is neglected~\cite{kramer2002macroscopic}.
We consider the general case of an anisotropic system described by the superfluid fraction tensor $\fs_{ij}$.
In this case, we have the modified Gross-Pitaevskii equation (GPE):
$$
i\hbar \partial_{t} \psi=-\frac{\hbar^{2}}{2m}\fs_{ij} \partial_{i} \partial_{j} \psi+V \psi+g|\psi|^{2} \psi
$$
where $\psi$ is the macroscopic wavefunction, $V$ is the external potential except the lattice (i.e., the harmonic trap in our experiment). The coupling constant $g=4\pi\hbar^2 a_s/m$ is proportional to the $s$-wave scattering length $a_s$.
Let $i,j=x,y,z$.
The mass current is defined as
$$J_{i}=\frac{\hbar}{2 i}\fs_{i j}\left(\psi^{*} \partial_{j} \psi-\psi \partial_{j} \psi^{*}\right).$$

By writing
$\psi=\rho e^{i \varphi}$, the GPE transforms into the hydrodynamics equations
$$
\left\{\begin{array}{l}
\partial_{t} \rho+\partial_{i} J_{i}=0 \rightarrow \partial_{t} \rho+ \frac\hbar{m}\partial_{i}\left(\rho\fs_{ij} \partial_{j} \varphi\right)=0 \\
-\partial_{t} \varphi=\frac{1}{2}\frac{\hbar}{m}\fs_{ij} \partial_{i} \varphi \partial_{j} \varphi+V+g \rho
\end{array}\right.
$$
From now on we consider the case that the lattice is along the principal axes, where $\fs$ is diagonal.

One can derive the angular momentum density for later use:
$$
\Pi_z = xJ_y-yJ_x=\frac{\hbar}{m}\fs_{xx} x\partial_y \varphi-\frac{\hbar}{m}\fs_{yy} y\partial_x \varphi.
$$

\subsection{Collective modes}
Now we can find the collective modes by making small perturbations about the equilibrium state, 
$$
\begin{aligned}
&\rho \rightarrow R+\delta \rho\\
&\varphi \rightarrow \varphi+\delta \varphi \\
&V \rightarrow V+\delta V
\end{aligned}
$$
$$
\left\{\begin{array}{l}
\partial_{t} \delta \rho+ \frac{\hbar}{m} \fs_{ i j}\left[\partial_{i} \delta \rho \partial_{j} \varphi+\partial_{i} \rho \partial_{j} \delta \varphi+\delta \rho \partial_{i} \partial_{j} \varphi+\rho \partial_{i} \partial_{j} \delta \varphi\right]=0 \\
-\partial_{t} \delta \varphi= \frac{\hbar^2}{m} \frac{1}{2}\fs_{i j}\left[\partial_{i} \varphi \partial_{j} \delta \varphi+\partial_{i} \delta \varphi\partial_{j} \varphi\right]+\delta V+g \delta \rho
\end{array}\right.
$$
Because of the initial state satisfies $\varphi=0$ and $\rho=\frac{\mu-\frac{1}{2} m \omega_{i}^{2} x_{i}^{2}}{g}$,
$$
\left\{\begin{array}{l}
\partial_{t} \delta \rho+ \frac{\hbar}{m} \fs_{ i j}\left[\partial_{i} \rho \partial_{j} \delta \varphi+\rho \partial_{i} \partial_{j} \delta \varphi\right]=0 \\
-\partial_{t} \delta \varphi=\delta V+g \delta \rho
\end{array}\right.
$$

Now we suppose 
$$
\delta \varphi=\delta \varphi_{x} x+\delta \varphi_{y} y
$$
to derive the dipole oscillation frequencies.
Let $\delta V=0$, 
$\delta \rho \rightarrow \delta \rho e^{-i\omega t}$ and
$\delta \varphi \rightarrow \delta \varphi e^{-i\omega t}$,
and collect coefficients before both linear terms $x$ and $y$,
one calculates the eigenmodes of
$$
\begin{pmatrix}
\omega^{2}-\fs_{xx} \omega_x^2 &  0 \\
0 & \omega^{2}-\fs_{yy} \omega_y^2
\end{pmatrix}
$$
The dipole mode frequencies are decoupled and are
$\omega_{x,d} = \sqrt{\fs_{xx}} \omega_x$ and 
$\omega_{y,d} = \sqrt{\fs_{yy}} \omega_y$.

The scissors mode is one of the quadratic modes, and we suppose
$$
\delta \varphi=\delta \varphi_{x x} x^{2}+\delta \varphi_{y y} y^{2}+\delta \varphi_{x y} x y
$$
and collect coefficients before all three quadratic terms (the calculation is omitted here), the resulting matrix is
$$
\begin{pmatrix}
\omega^{2}-3 \omega_{x}^{2} \fs_{x x} &  \quad-\omega_{x}^{2} \fs_{y y} & -2  \omega_{x}^{2}\fs_{x y} \\
-\fs_{x x}  \omega_{y}^{2} &  \quad \omega^{2}-3 \omega_{y}^{2} \fs_{y y} & -2  \omega_{y}^{2}\fs_{x y}
\right.\\
-2 \omega_{y}^{2}\fs_{x y} &  \quad-2  \omega_{x}^{2}\fs_{x y} &  \quad \omega^{2}-\left(\fs_{x x} \omega_{x}^{2}+\fs_{y y}\omega_{y}^{2}\right)
\end{pmatrix}
$$
If $\fs_{x y}=0$, $\fs_{x x}\neq \fs_{y y}$,
$$
\left(\begin{array}{ccc}
\omega^{2}-3 \fs_{xx} \omega_{x}^{2} & -\fs_{yy} \omega_{x}^{2} &  0 \\
-\fs_{xx} \omega_{y}^{2} & \omega^{2}-3 \fs_{yy} \omega_{y}^{2} & 0 \\
0 & 0 & \omega^{2}-\fs_{xx}\omega_{x}^{2}-\fs_{yy}\omega_{y}^{2}
\end{array}\right)
$$
The scissors mode frequency is $\omega_{sc}=\sqrt{\fs_{xx}\omega_{x}^{2}+\fs_{yy}\omega_{y}^{2}}$, while
the two quadruple mode frequencies are
$$
\omega_{xx,yy}=
\sqrt{ \frac{3 \bar{\omega}^{2} \pm \sqrt{9 \bar{\omega}^{4}-32 \fs_{xx} \fs_{yy} \omega_x^{2} \omega_{y}^{2}}}{2} }
$$
with 
$\bar{\omega}=\omega_{sc}$.

It can be easily seen from the wavefunction that the scissor mode is decoupled from the other two quadrupole modes. Actually this can be similarly generalized to 3D, where one has three different scissors mode $xy, yz, zx$. Therefore, although our system is actually 3D, the scissors mode frequency is not changed from the simple 2D result derived here.

An additional note is that this procedure can be easily extended to the case of $\fs_{xy}\neq 0$, which can be made experimentally by letting the lattice direction deviate from both trap axes.

\subsection{Scissors rotation}
Following the above section
$$
\Pi_z = xJ_y-yJ_x=\frac{\hbar}{m}\fs_{xx} x\partial_y \varphi-\frac{\hbar}{m}\fs_{yy} y\partial_x \varphi,
$$
and plug in the second hydrodynamics equation
\begin{equation}
\label{eq:hydro_Pi_z}
    \partial_t \Pi_z = x\partial_t J_y-y\partial_t J_x=-x\partial_j T_{jy}+y\partial_j T_{jx}-x\rho\partial_y(\fs_{yy}\frac{V}{m})+y\rho\partial_x(\fs_{xx}\frac{V}{m}).
\end{equation}
where we define 
$T_{jk} := \rho v_jv_k-\sigma_{jk}$ and
$\sigma_{jk} :=-\frac12\delta_{jk}g(\frac{\rho}{m})^2+(\frac{\hbar}{2m})^2\rho\partial_j\partial_k\ln{\rho}$.
Those are actually the counterparts of the classical momentum flux and stress tensor.
We did not neglect the quantum pressure here.

The integration over the space generates the torque 
$$
\tau = \int dx dy ( \fs_{xx}\omega_x^2-\fs_{yy}\omega_y^2)\rho xy
=( \fs_{xx}\omega_x^2-\fs_{yy}\omega_y^2)\langle xy \rangle
$$
where the first two terms in the equation~\ref{eq:hydro_Pi_z} vanish because they are total derivatives.
Consider the cloud being rotated by a small angle $\theta$, the torque is derived as
\begin{equation}
    \tau = ( \fs_{xx}\omega_x^2-\fs_{yy}\omega_y^2)\langle x'^2-y'^2 \rangle \theta
\end{equation}
where $\langle x'^2-y'^2 \rangle$ depends only on the initial Thomas-Fermi distribution of the cloud, as $\theta\rightarrow0$.

Now we have an equation of motion similar to that of a pendulum in classical physics,
$$
\tau = \partial_t{L_z} := I\ddot{\theta}
$$
where we define $I$ as the moment of inertia of the gas.
The moment of inertia $I$ then can be measured by measuring the oscillation frequency $\omega_{sc}$ in the experiment:
$$
\frac{I}{I_c} = \frac{-\tau/\theta }{I_c\omega_{sc}^2} = \frac{ ( \fs_{xx}\omega_x^2-\fs_{yy}\omega_y^2)\langle x^2-y^2 \rangle }{\langle x^2+y^2 \rangle ~\omega_{sc}^2}
$$
where $I_c$ is moment of inertia a classical mass distribution
$I_c=\langle x^2+y^2 \rangle$.
One notes that the expression is different from that derived from the simple Lagrangian or the angular momentum sum-rule naively. 

\subsection{Moment of inertia}
The moment of inertia can also be derived its definition 
$$
I
    = \frac{\partial \braket{L_z}}{\partial \Omega}.
$$
We calculate this derivative respectively in the cases of (ii) static lattice and (ii) rotating lattice.

The main difference between case (i) and (ii) is the order of two operations: 
the projection into the lowest band of the lattice and rotating frame transformation that makes the trap potential time invariant. 
In a rotating trap but static lattice, we first project the dynamics to the lowest band, and then the time dependent potential can be transformed away and one derives additional terms with the angular frequency $\Omega$,
\begin{equation}
\label{eq:eff_mass_hydro_stat}
    \left\{\begin{array}{l}
    \partial_{i}\left(\frac{\hbar}{m}\rho\fs_{i j} \partial_{j} \varphi\right)-\nabla \cdot(\rho \vec{\Omega} \times \vec{r})=0 \\
    \frac{\hbar^2}{2m^2}\fs_{i j} \partial_{i} \varphi \partial_{j} \varphi+\frac{V}{m}+\frac{g \rho}{m}- \frac{\hbar}{m}\nabla\varphi \cdot(\vec{\Omega} \times \vec{r})=\frac{\mu}{m}
    \end{array}\right.
\end{equation}
Here we use the ansatz $\varphi=\frac{m}{\hbar}\alpha x y$. Since the $\mu$ in the second equation is spatial independent,  and under the $\Omega\rightarrow0$ limit, we obtain
$$
\frac{y\partial_x\rho}{x\partial_y\rho}=\frac{\omega_x^2}{\omega_y^2},
$$
where the terms of $\alpha$ and $\Omega$ are neglected.
The first equation gives
$$
\frac{y\partial_x\rho}{x\partial_y\rho}=\frac{\Omega-\fs_{yy}\alpha}{\Omega+\fs_{xx}\alpha}
.$$
Therefore we obtain
\begin{equation}
    \label{eq: alpha}
    \alpha=-\Omega \frac{\omega_{x}^{2}-\omega_{y}^{2}}{\fs_{xx} \omega_{x}^{2}+\fs_{yy}\omega_{y}^{2}},
\end{equation}
and the total angular momentum becomes
$$\begin{aligned}
\left<L_z\right>&=\left<\vec{r}\times \vec{J}\right>=\alpha\left<\fs_{yy}x^2-\fs_{xx}y^2\right>=\frac{\omega_{x}^{2}-\omega_{y}^{2}}{\fs_{xx} \omega_{x}^{2}+\fs_{yy}\omega_{y}^{2}}\left<\fs_{xx}y^2-\fs_{yy}x^2\right>\Omega
\end{aligned}
$$

\begin{equation}
\label{eq:static_I}
    \frac{I}{I_c} 
    = \frac{1}{I_c} \frac{\partial \braket{L_z}}{\partial \Omega} = 
    \frac{\omega_x^2-\omega_y^2}{\omega_x^2+\omega_y^2} \frac{\fs_{xx}\omega_x^2-\fs_{yy}\omega_y^2}{\fs_{xx}\omega_x^2+\fs_{yy}\omega_y^2} 
\end{equation}
where $I_c$ is moment of inertia a classical mass distribution
$I_c=\langle x^2+y^2 \rangle$.
One can verify that the scissors mode frequency is given by
\begin{equation}
    \omega_{sc}=\sqrt{\frac{-\tau}{I\theta}}=\sqrt{\fs_{xx}\omega_x^2+\fs_{yy}\omega_y^2}
\end{equation}
the same as derived from the collective mode section. 

The most striking observation from the result is that the moment of inertia can go negative when the superfluid density along one axis is suppressed below a critical value determined by the trap frequencies. 
This behavior is purely quantum mechanical, and has no counterpart in classical physics as we know of.
From a hydrodynamics view, the reason for this zero-crossing is that the angular momentum density always has co-rotating and counter-rotating parts in a quantum gas, due to the irrotational nature of the order parameter.
Without the lattice, however, the co-rotating part always exceeds the counter-rotating part, thus leading to positive moment of inertia. 
But with the anisotropic lattice present the relative contribution of them can be re-tuned by varying the lattice depth. 

Next we consider the case (ii) with a rotating lattice synchronized with the rotating harmonic trap. In this case, the projection operation into the lowest band needs to be taken after the rotating frame transformation. The equation~(\ref{eq:eff_mass_hydro_stat}) then needs to modified as
\begin{equation}
\label{eq:eff_mass_hydro_rot}
    \left\{\begin{array}{l}
    \partial_{i}\left(\frac{\hbar}{m}\rho\fs_{i j} \partial_{j} \phi\right)-\nabla \cdot(\rho \vec{\tilde\Omega} \times \vec{r})=0 \\
    \frac{\hbar^2}{2m^2}\fs_{i j} \partial_{i} \phi \partial_{j} \phi+\frac{V}{m}+\frac{g \rho}{m}- \frac{\hbar}{m}\nabla\phi \cdot(\vec{\tilde\Omega} \times \vec{r})=\frac{\mu}{m}
    \end{array}\right.
\end{equation}
where the 
$\tilde{\Omega}$
is selected to be
$\tilde{\Omega}=\Omega(\fs_{xx},\fs_{yy})$.
This can be seen from the fact that the kinetic energy part is modified by the superfluid fraction tensor, so is the current operator.
One arrives at similar to the equation~(\ref{eq: alpha})
\begin{equation}
    \alpha=-\Omega \frac{\fs_{xx}\omega_{x}^{2}-\fs_{yy}\omega_{y}^{2}}{\fs_{xx} \omega_{x}^{2}+\fs_{yy}\omega_{y}^{2}}
\end{equation}
and
\begin{equation}
\label{eq:sf_I}
    \frac{I^{sf}}{I_c}=
    \frac{(\fs_{xx}\omega_x^2-\fs_{yy}\omega_y^2)^2}{(\fs_{xx}\omega_x^2+\fs_{yy}\omega_y^2)(\omega_x^2+\omega_y^2)} 
\end{equation}
This is the superfluid contribution to the moment of inertia, because only the superfluid flow can be derived from the coarse-graining process that assumes lowest band dynamics. Equation~(\ref{eq:sf_I}) aligns very well with the simulation result which is calculated from the gradients of the coarse-grained superfluid phase $\vartheta(\mathbf{r})$, confirming the self-consistent superfluid description.
Note that it is strictly positive and rather different from the equation~(\ref{eq:static_I}), due to the different order of rotating frame transformation and coarse graining.

However, the rest part of the moment of inertia is contributed by the normal fluid indeed. The normal fluid current can be written as
\begin{equation}
    J^n= (-\rho^n_{xx} y, \rho^n_{yy} x)\Omega
\end{equation}
which can be derived from the transverse current definition
\begin{align}
\label{eq:transverse_curr}
I = \lim_{\mathbf{q}\to 0}\sum_{n \neq 0} \frac{|\bra{0} \hat L_z e^{i \mathbf{q}\cdot \mathbf{r}} \ket{n}|^2 + |\bra{0} 
\hat L_z e^{-i \mathbf{q}\cdot \mathbf{r}} \ket{n}|^2}{E_n-E_0},
\end{align}
and it gives rise to
\begin{equation}
    \frac{I^{n}}{I_c}=
    \frac{\fn_{xx} \omega_x^2 +\fn_{yy} \omega_y^2 }{\omega_x^2+\omega_y^2}
\end{equation}
This result is also checked with the simulation by calculating the subtraction of the superfluid from the total moment of inertia.

\bibliography{main}